\documentstyle[twocolumn,aps]{revtex}
\input{epsf}
\begin{document}
\draft

\title{Intermittency and turbulence in a magnetically confined fusion plasma}
\author{V. Carbone$^1$, L. Sorriso--Valvo$^1$, E. Martines$^2$, 
V. Antoni$^{2,3}$, and P. Veltri$^1$}

\address{$^1$ Dipartimento di Fisica, Universit\`a degli studi della Calabria,
87036 Rende (CS), Italy, \\ and Istituto Nazionale di Fisica della
Materia Unit\`a di Cosenza \\
$^2$ Consorzio RFX, corso Stati Uniti 4, 35127 Padova, Italy\\
$^3$ Istituto Nazionale di Fisica della Materia, Unit\`a di Padova}
\date{\today}
\maketitle

\begin{abstract}

We investigate the intermittency of magnetic turbulence as measured in 
Reversed Field Pinch plasmas. We show that the Probability Distribution 
Functions of magnetic field differences are not scale invariant, that is the 
wings of these functions are more important at the smallest scales, a 
classical signature of intermittency. We show that scaling laws appear also in 
a region very close to the external wall of the confinement device, and we 
present evidences that the observed intermittency increases moving towards the 
wall.

\end{abstract}


%\begin{multicols}{2}

The issue of self--similarity is of paramount importance in turbulence studies.
Indeed, self--similarity is one of the key hypotheses of Kolmogorov theory
\cite{Kolmogorov,Frisch}, which leads for fluid turbulence to the famous
$-5/3$ exponent for the power spectrum decay in the inertial range (the
intermediate range of scales between the large scales where energy is injected
and the small ones where it is dissipated). Notwithstanding the success of the
Kolmogorov's theory, the study of the Probability Distribution Functions (PDF)
of velocity fluctuations at a given scale has shown a departure from
gaussianity  in the PDF tails. The same phenomenon is usually
evidenced by looking at the scaling exponents of higher order moments of
fluctuations, which appear to be nonlinear functions of the order index.
Intermittency described by multifractal models \cite{Frisch} is usually
invoked to be the cause of the observed break of pure self--similarity. Even
if Kolmogorov theory was  originally developed for homogeneous and isotropic
turbulence, the evidence for self--affine fields has been studied also inside
the boundary layer  turbulence in fluid laboratory experiments \cite{Benzi99}
which is neither  homogeneous nor isotropic. The only hypothesis required to
perform these  studies from an experimental point of view, in order to apply
the usual  Taylor's hypothesis, is the statistical stationarity of turbulence.
Recently it has been shown \cite{so3} that only after a suitable decomposition
in terms of irreducible representations of the $SO(3)$ groups one can hope to
properly disentangle isotropic from anisotropic effects in Navier--Stokes
equations. Of course this should be true also in MHD flows, even if it is not
clear how to recover anisotropic and non homogeneous effects from real data.  

While in ordinary fluids the statistical properties of turbulence have been
well characterized, both theoretically and experimentally, in magnetized
fluids only recently this has been undertaken, mostly in relation with
velocity and magnetic field fluctuations measured in the solar wind
\cite{Solarwind}. In this paper we report evidences for
the presence of intermittency in another type of magnetized fluid, namely a
plasma of interest for controlled thermonuclear fusion research, confined in
reversed field pinch (RFP) configuration.

The RFP is a configuration of magnetic fields \cite{Bodin}
characterized by toroidal and
poloidal components of comparable magnitude (in a tokamak the
field is mainly toroidal). 
The configuration represents a near-minimum energy state to which a plasma
relaxes under proper constraints \cite{Taylor}.
The toroidal field changes sign in the outer part of the plasma, a feature
which gives the name to the configuration. Such field reversal, which
improves the MHD stability of the configuration, is spontaneously generated
by the plasma, and is maintained against resistive diffusion by the dynamo
process \cite{Biskamp}.
This is achieved through the action and nonlinear coupling of
several resistive magnetohydrodynamic (MHD) modes, which give rise to a high
level of magnetic turbulence (of the order of 1\% of the average field in
present day experiments, i.e. two orders of magnitude larger than in
tokamaks). This high fluctuation level makes the RFP very suited
for the study of MHD turbulence properties, mainly for their magnetic part.
The magnetic turbulence has been demonstrated to be the main cause of
energy and particle transport in the RFP core, whereas at the edge
its contribution is still under investigation. In this region the
electrostatic turbulence has been proved to give an important
contribution to the particle transport \cite{Antoni98}. It is worth
mentioning that a recent investigation of the edge electrostatic
turbulence in different fusion devices (including RFP and tokamaks) has
shown the existence of long range time and space correlations
\cite{Carreras}.

The measurements described in this paper have been obtained in the RFX
experiment, which is the largest RFP presently in operation (major radius $2$
m, minor radius $0.457$ m) \cite{Fellin}. RFX is designed to reach a plasma
current of $2$ MA, and currents up to $1$ MA have been obtained up to now. The
measurements were performed in low currents discharges ($300$ kA) using a
magnetic probe inserted in the edge plasma. The probe consists of a coil
housed in a boron nitride protecting head. The coil measures the time
derivative $\partial_t B$ of the radial component $B(t)$ of the magnetic
field. The radial direction in this case goes from the core plasma to the
edge. The sampling frequency of the measurements is $2$ MHz. Measurements have 
been collected at different values of the normalized radius $r/a$ ($r/a=1$
identifies the location of the last magnetic flux surface)
%the 
%$r/a = 1$ surface is identified as that where the floating potential measured 
%with respect to the first wall vanishes). 
In RFX two different components of
the magnetic fluctuations can be identified: a localised and stationary
magnetic perturbation, originated by the tearing modes responsible for the
dynamo which tend to be phase--locked and locked to the wall \cite{locking},
and an high frequency broadband activity, which is investigated here.
All measurements presented were made away from the stationary perturbation.

\begin{figure}[htb]
\epsfxsize=9cm    % dimensioni della figura
\centerline{\epsffile{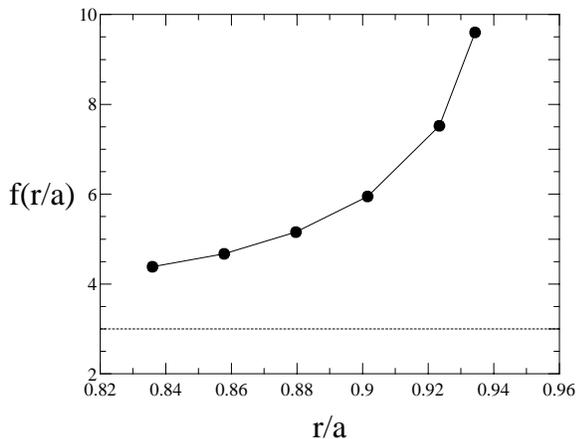}}
\caption{We show the values of the flatness $f(r/a)$ for the derivative of the
radial magnetic field, as a function of the insertion $r/a$. The gaussian
value ($f=3$) is shown as a dotted line.}
\label{figdery}
\end{figure}

We start by looking at the statistical properties of the normalized variables
$s(t) = \partial_t B/\sqrt{<(\partial_t B)^2>}$ (brackets being time
averages). In figure \ref{figdery} we show the flatness of these stochastic
variables $f = <s^4>/[<s^2>]^2$ for some positions $r/a$. As it can be seen
$f(r/a)$ is higher than the gaussian value and tends to decrease as $r/a$
increases. This is a first rough evidence that the observed
magnetic field is intermittent, that is the time evolution of $\partial_t B$
is dominated by
strong magnetic fluctuations. The intermittency (say the departure from a
gaussian statistics) is more visible near the external wall.

To get some insight into the nature of intermittency actually present in the
fusion device, following the usual analysis currently made in fluid flows
\cite{Frisch}, we investigate the scaling behavior of the stochastic variables
$\delta B(\tau) = B(t+\tau)-B(t)$, which represents characteristic fluctuations
across turbulent structures at the scale $\tau$. For each position within the
device, we can study the statistical behavior of fluctuations at different
scales $\tau$. The interest of this resides in the fact that, if we introduce a
scaling law for magnetic fluctuations $\delta B(\tau) \sim \tau^h$ as MHD
equations seem to indicate \cite{Carbone96}, a scale variation $\tau \to
\lambda \tau$ ($\lambda$ is the parameter which defines the change of scale)
leads to

\[
\delta B(\lambda \tau) = \lambda^{-h} \delta B(\tau)
\]
This is interpreted as an ``equality in law'' \cite{Frisch}, that is the
right--hand--side of the equation has the same statistical properties
of left--hand--side. If $h$ is constant we can easily show that the PDFs of 
the normalized stochastic variables
$\delta b_{\tau} = \delta B(\tau)/\sqrt{<[\delta B(\tau)]^2>}$ collapse to 
a unique PDF, independent on the scale $\tau$. This is true in a pure
self--similar (fractal) case. On the contrary we must invoke the multifractal
model to describe intermittency \cite{Frisch} which is introduced by defining
a range of values of $h$.

\begin{figure}[htb]
\epsfxsize=10cm    % dimensioni della figura
\centerline{\epsffile{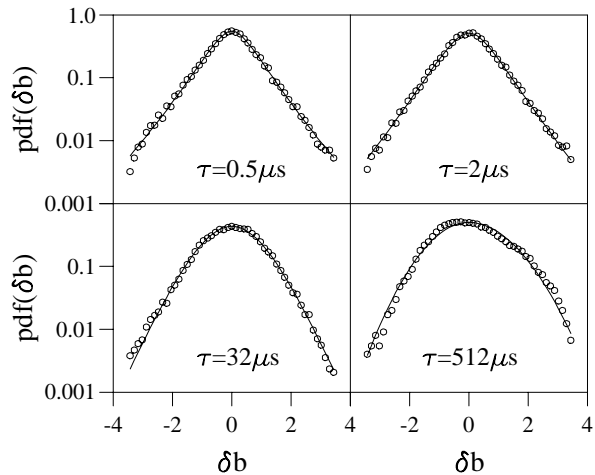}}
\caption{We show the PDFs of the normalized magnetic fluctuations for four 
different scales, at a given position $r/a = 0.95$. The full line represents 
the fit made with the convolution function.}
\label{figpdfs}
\end{figure}

In figure \ref{figpdfs} we report the PDFs of $\delta b_{\tau}$ at different
scales for a given value of $r/a$. As it can be seen the PDFs do not collapse
to a single curve, but follow a characteristic scaling behavior which is
visible for all values of $r/a$. At large scales the PDF are almost gaussian,
and the wings of the distributions grow up as the scale becomes smaller.
Stronger events at small scales have a probability of occurrence greater than
that they would have if they were distributed according to a gaussian
function. This behavior, that is the presence of self--affine fields, is at the
heart of the phenomenon of intermittency as currently observed in fluid flows
\cite{Frisch,Castaing}.

The behavior of PDFs against the scale can be described by introducing a
given shape for the distribution. At each scale $\tau$ the PDF of
$\delta b_{\tau}$ can be represented as a convolution of gaussian functions of
widths $\sigma$ whose distribution is given by a function $G_{\lambda}(\sigma)$

\begin{equation}
P(\delta b_{\tau}) = {1 \over \sqrt{2 \pi}} \int_0^{\infty}
G_{\lambda}(\sigma) \exp \left (-\delta b_{\tau}^2 / 2 \sigma^2 \right)
{d\sigma \over \sigma}
\label{pdfs}
\end{equation}
which can be interpreted in the framework of a cascade model as the signature
of an
underlying multiplicative process \cite{Castaing,Vulpio,Sorriso}. We use a
log--normal ansatz

\begin{equation}
G_{\lambda}(\sigma) = {1 \over \sqrt{2 \pi} \lambda}
\exp\left(-{\ln^2 \sigma/\sigma_0 \over 2 \lambda^2}
\right)
\label{lognormal}
\end{equation}
even if other functions does not give really different results 
\cite{Castaing}. The free parameter $\lambda^2$ represents the width of the
distribution $G_{\lambda}$, while $\sigma_0$ is the most probable value
of $\sigma$. The scaling behavior of $P(\delta b_{\tau})$ is
translated in the scaling variation of the parameter $\lambda^2$ 
\cite{Castaing,Sorriso}. In fact when the PDF is gaussian $\lambda^2 = 0$ 
($G_{\lambda}$ becomes a delta function centered around $\sigma_0$), while 
the departure from a gaussian function increases as $\lambda^2$ increases. 
In figure \ref{figpdfs} we report as full line a fit of the data with
equation (\ref{pdfs}). A satisfactory agreement at all scales is evident.

\begin{figure}[htb]
\epsfxsize=10cm    % dimensioni della figura
\centerline{\epsffile{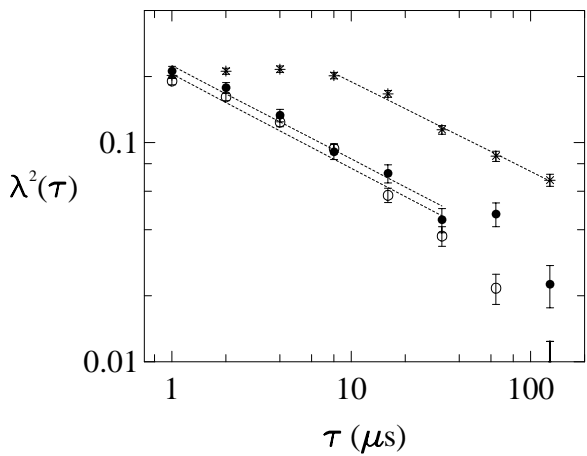}}
\caption{Scaling behavior of the exponent $\lambda^2(\tau,r/a)$ for three 
different insertion $r/a$, namely: $r/a = 0.97$ (black circles), $r/a = 0.95$ 
(white circles), and $r/a = 0.91$ (stars).}
\label{figparafit}
\end{figure}

Looking at the scaling laws for $\lambda^2$, at different insertion
points $r/a$, it can be seen (figure \ref{figparafit}) that $\lambda^2$
displays a power law behavior
\[
\lambda^2(\tau,r/a) = \Lambda^2(r/a) \tau^{\beta}
\]
all over the observed time scales, for insertion points
near the external wall. On the contrary measurements more inside the device
show a saturation of intermittency at scales $\tau_S \simeq 10$ $\mu$s. The
values of $\beta$ we find are of the order of $\beta \simeq 0.42 \pm 0.03$,
close to that found for the velocity field both in fluid flows and in the 
solar wind turbulence \cite{Castaing,Sorriso}, but higher than the value 
found for the magnetic field intensity in the solar wind \cite{Sorriso}. 
Finally the absolute values for $\lambda^2$ are 
decreasing going from the wall inside the device. Namely we found
$\lambda^2_{max} = 0.21 \pm 0.01$ for $r/a = 0.98$, and $\lambda^2_{max} = 
0.086 \pm 0.006$ for $r/a = 0.84$.
%(not displaied on figure \ref{figparafit}).
This is a further confirmation of the stronger intermittency near the 
external wall. All the error--bars has been estimated starting from a 
poisson statistical uncertainity on the PDFs value.

\begin{figure}[htb]
\epsfxsize=10cm    % dimensioni della figura
\centerline{\epsffile{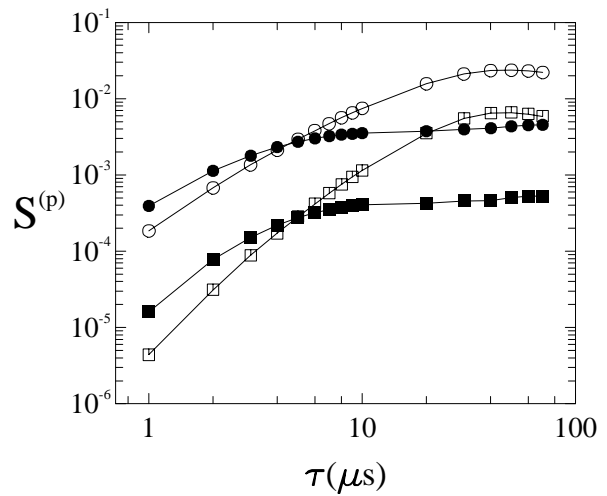}}
\caption{The structure functions $S_{\tau}^{(p)}$ are shown for $p = 2$ 
(circles) and $p = 3$ (squares). Open symbols refer to the position 
$r/a = 0.96$, full symbols refer to $r/a = 0.86$.}
\label{figsf}
\end{figure}

A complementary analysis of intermittency can be performed by calculating
the scaling exponents of the structure functions, say of the $p$--th moments
of fluctuations $S_{\tau}^{(p)} =
< \delta b_{\tau}^p >$ (brackets are defined as time averages). In figure 
\ref{figsf} we report the structure functions $S_{\tau}^{(p)}$, for two 
values of $p$, and for two different position $r/a$. The differences for 
different position is evident, and represent a signature of the absence of
universality.

To calculate the scaling exponents, we use the generalized scaling introduced 
by Benzi and coworkers \cite{Benzi}, which has been found to be useful also in 
magnetohydrodynamic turbulence \cite{Carbone96,mhdess}, thus obtaining the 
normalized scaling exponents $\zeta_p$ defined through 
$S_{\tau}^{(p)} \sim [S_{\tau}^{(3)}]^{\zeta_p}$. 

\begin{figure}[htb]
\epsfxsize=9cm    % dimensioni della figura
\centerline{\epsffile{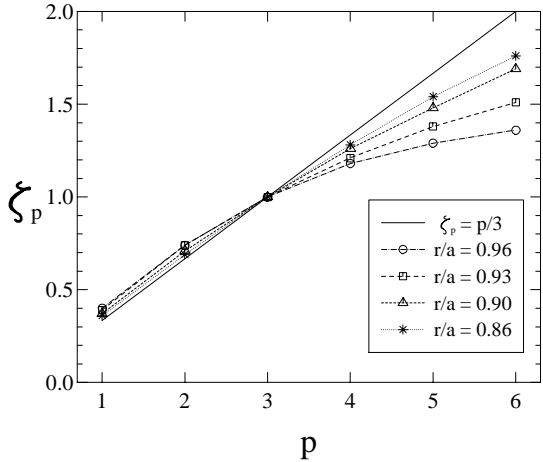}}
\caption{The normalized scaling exponents $\zeta_p$ of the structure functions
are shown as a function of $p$, for different insertion points $r/a$.
Errorbars, about $5\%$ of the exponent values, are not displayed for clarity.  
The K41 scaling $\zeta_p \sim p/3$ is also reported for comparison.}
\label{figexpo}
\end{figure}

In figure \ref{figexpo} we report the
scaling exponents obtained for some insertions $r/a$. The behavior of $\zeta_p$
against $p$ shows that scaling exponents are anomalous, say they are different
from the usual $p/3$ Kolmogorov's law. Note once more that the strength of
intermittency, measured through the difference between $\zeta_p$ and $p/3$,
is greater near the wall.
In conclusion scaling laws for PDFs of magnetic fluctuations, and anomalous 
scalings for structure functions, are found everywhere in the outer plasma 
region of the RFX thermonuclear fusion experiment.
We find that
the anomaly of scaling exponents, as well as scaling laws for PDFs, strongly
depend on the position inside the plasma, so that magnetic turbulence inside 
the device is not universal, as far as scaling laws are concerned. 
Possible reasons for this are the presence of
the first wall, the presence of the toroidal field reversal (which takes place 
at $r/a \simeq 0.9$) or the strongly sheared plasma flow measured in 
the RFX edge \cite{Antoni97}. Concerning this latter option, it is 
worth to mention that in principle different plasma velocities in 
different points would only affect the relationship between time and 
spatial scales obtained through Taylor hypothesis, and not the PDF 
scaling properties. However, the eddy breaking effect induced by a velocity 
shear is well known to affect electrostatic turbulence in fusion 
plasmas\cite{Burrell,Antoni98}, and an influence on MHD turbulence can  also be
envisaged, either directly or thorugh nonlinear coupling to electrostatic modes.
If this is not the case, the reason for the
observed differences could be perhaps found in the conjecture of Farge
\cite{Farge}. She proposed that turbulence could be described by
interwoven sets of both intermittent structures and background
gaussian flow on each characteristic scale. The nature of the intermittent
structures can evidently be influenced by walls \cite{Benzi99}, and/or
current sheets associated with field reversal \cite{Veltri}.
We are actually reviewing and testing this idea
on the RFX device in order to identify structures which generate
intermittency. Since a reduction of magnetic fluctuations has been linked
to improvements in the energy confinement \cite{Bartiromo}, a
better understanding of the generation of intermittency through structures
could improve the confinement physics understanding.

\acknowledgments{We are grateful to Francesco Pegoraro for some discussions 
and for its interest in this work.}

\end{document}